# Conversion between metavalent and covalent bond in metastable superlattices composed of 2D and 3D sublayers

Dasol Kim[1,2,▽], Youngsam Kim[3,▽], Jin-Su Oh[4,▽], Changwoo Lee[1,▽], Hyeonwook Lim[1,▽], Cheol-Woong Yang[4,*], Eunji Sim[3,*] and Mann-Ho Cho[1,5,*]

[1]Department of Physics, Yonsei University, Seoul, Republic of Korea.

[2]Presently at Aachen. I. Institute of Physics, Physics of Novel Materials, RWTH Aachen University, 52056 Aachen, Germany

[3]Department of Chemistry, Yonsei University, Seoul, Republic of Korea.

[4]School of Advanced Materials Science and Engineering, Sungkyunkwan University, Suwon, Republic of Korea.

[5]Department of System Semiconductor Engineering, Yonsei University, Seoul, Republic of Korea.

* corresponding author: cwyang@skku.edu, esim@yonsei.ac.kr, and mh.cho@yonsei.ac.kr

**Abstract**

Reversible conversion over multi-million-times in bond types between metavalent and covalent bonds becomes one of the most promising bases for universal memory. As the conversions have been found in metastable states, extended category of crystal structures from stable states via redistribution of vacancies, researches on kinetic behavior of the vacancies are highly on demand. However, they remain lacking due to difficulties with experimental analysis. Herein, the direct observation of the evolution of chemical states of vacancies clarifies the behavior by combining analysis on charge density distribution, electrical conductivity, and crystal structures. Site-switching of vacancies gradually occurs with diverged energy barriers owing to a unique activation code—the accumulation of vacancies triggers spontaneous gliding along atomic planes to relieve electrostatic repulsion. Study on the behavior can be further applied to multi-phase superlattices composed of $Sb_2Te_3$ (2D) and GeTe (3D) sublayers, which represent the superior memory performances but their operating mechanisms were still under debates due to their complexity. The site-switching is favorable (suppressed) when Te-Te bonds are formed as physisorption (chemisorption) over the interface between $Sb_2Te_3$ (2D) and GeTe (3D) sublayers driven by configurational entropic gain (electrostatic enthalpic loss). Depending on the type of interfaces between sublayers, phases of the superlattices are classified into metastable and stable states, where the conversion could be only achieved in the metastable state. From this comprehensive understanding on the operating mechanism via kinetic behaviors of vacancies and the metastability, further studies towards vacancy engineering are expected in versatile materials.

Metavalent bond (MVB) is a recently defined new type of bond to clarify exotic properties of tons of condensed materials which cannot be rationalized by neither covalent bond (CVB) nor metallic bond. MVB successfully develops design rules for functional materials such as phase-change memory (PCM), thermoelectric, photo-voltaic, and battery materials. In the case of PCM, conversion in bond type between MVB and CVB with drastic contrast in electrical resistances is utilized as a basis to store data, which becomes one of the most promising candidate for universal memories, photonic computing, and neuromorphic hardware.[1–7] Their superior memory performances have been found in metastable states rather than stable states of chalcogenide PCMs such as ultrafast, energy-efficient, and reversible conversion over multi-million-times. Various studies reported randomly distributed vacancies in metastable states as an origin of superior memory performances that vacancies provide structural flexibility to repeat the conversion without severe structural deformation or phase-separation.[8,9]

Vacancies are inherently generated in chalcogenide PCMs to localize excess electrons occupying atomic sites next to chalcogen as the number of valence electrons of chalcogen ($p^4$) exceeds that of orbital configuration of crystal structures ($p^3$). It is worth to note that formation of vacancies is suppressed in pnictogenide PCMs because the number of valence electrons of pnictogen ($p^3$) corresponds to that of orbital configuration of crystal structures ($p^3$). The study on conversion in bond type in metastable chalcogenide PCMs with vacancies triggered developments of enhanced materials, i.e., superlattices composed of $Sb_2Te_3$ (2D) and GeTe (3D). This system has received considerable scientific attention owing to its unique interfaces between the 2D and 3D sublayers and technical attention owing to it having the highest memory performances. [10–12] However, their operating mechanisms are still inconclusive over a decade due to structural complexity and difficulty to experimentally analyze vacancies. Further, in the case of superlattices, conversion in bond types and meta-stability have been yet clarified.

Herein, based on the direct observation in the evolution of the chemical states of vacancies, the kinetic behaviors of vacancies were studied in metastable and stable $Sb_2Te_3$, as a prototype of chalcogenide PCMs. The site-switching of vacancies gradually occurs with diverged energy barriers through gliding along atomic planes that the kinetics of the vacancies can be considered as a second-order transition.[13,14] The behavior is extended to $Sb_2Te_3$/GeTe superlattices that the site-switching of vacancies is suppressed or encouraged depending on the types of interfacial structure, where combination of those interfaces classifies the stability of the phases into metastable and stable states. Notably, conversion in bond types between the metavalent bond (MVB) and covalent bond (CVB) are corroborated in metastable superlattices with transition of local environments of Ge. [15–18] We believe that this study provides a novel perspective of the kinetics in the vacancies, where the metastable and stable states are expected to be applicable to other functional materials with vacancies.

**Results and Discussion**

Quintuple layers (Te–Sb–Te–Sb–Te, QLs) are stacked over a van der Waals (vdW) layer repeatedly in hexagonal $Sb_2Te_3$, as displayed in the left panel of Figure 1a and Figure S1 in the Supporting Information. The stacking order over the vdW layer between the QLs in $Sb_2Te_3$ is relatively well preserved in comparison to that of other compounds such as magic-angle graphenes, which agrees with subsequent reports in that there are many electrons at the vdW layers of $Sb_2Te_3$ compared with the other vdW compounds such as mono-chalcogenides (GaSe). [19–22] In further detail, we corroborate the spatial distribution of the excess electrons at the vdW layers using charge density differences (CDDs, dashed orange box). Interestingly, those are on (off) the alignment of the intraQL (interQL) to minimize the electrostatic repulsion. Excess p-orbital electrons near the vdW layer from the upper QL are aligned toward the $a$-site, whereas those from the lower QL are aligned toward the $c$-site. Here, the $a$, $b$, and $c$- sites represent the stacking sequence in the plane of the atomic plane perpendicular to the vdW layer without loss of generality. Therefore, the excess electrons leave the b-site relatively vacant, which can be identified by plotting the charge densities (CDs) on the isosurface (grey) at the lowest level of density (0.003 e bohr$^{-3}$); the isosurfaces at various levels of density are shown in Figure S2 in the Supporting Information. The periodically localized sites (red balls) on the vdW layers seem to occupy atomic sites, where they can be regarded as "inherent vacancies". Considering vacancies as additional atomic sites (vac$_1$Sb$_2$Te$_3$), the periodically localized sites on the vdW layer can be well explained by the electron counting model: the excess *p*-electrons of Sb$_2$Te$_3$ are relaxed from 3.6 to 3 on average, which corresponds to the *p$^3$*-orbital configuration; further discussions are provided in Figure S3 in the Supporting Information. It is worth noting that the relaxation typically occurs in VA-VIA compounds, where the most stable structures in the composition are sesqui-chalcogenides with vdW layers (VA$_2$VIA$_3$ or vdw$_1$VA$_2$VIA$_3$), such as Bi$_2$Se$_3$ and Bi$_2$Te$_3$. In other words, inherent vacancies relieve the electrostatic repulsion by misaligning the excess electrons over the vdW layer, lowering the enthalpy. Therefore, an inherent vacancy is differentiated from the vacancy defects, which are established in the gain of configurational entropy. As presented in Table 1, the density of the inherent vacancies is nearly fixed and comparable to that of the compositional elements (~10$^{22}$ atoms cm$^{-3}$), whereas that of vacancy defects is adjustable and considerably lower (<< 10$^{20}$ atoms cm$^{-3}$) than that of the elements.

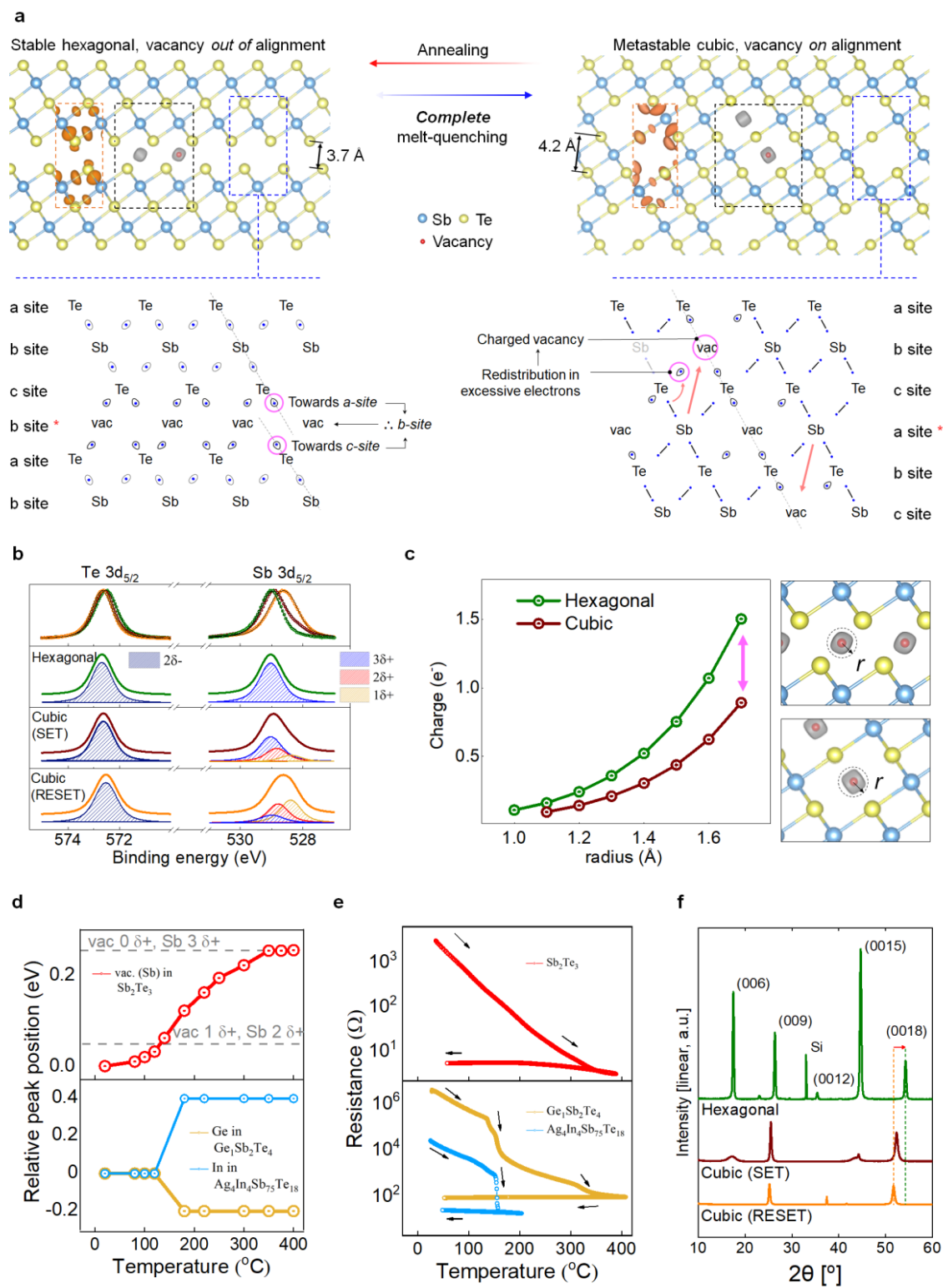

**Figure 1. Phase transition of Sb$_2$Te$_3$.** (a) Hexagonal and cubic Sb$_2$Te$_3$. (b) Chemical states of Te and Sb depending on the crystallinity. (c) Chemical states at vacancy sites obtained by DFT calculations. (d, e, f) On annealing, evolution in the (d) chemical states, (e) electrical resistance, and (f) structures.

**Table 1.** Comparison between inherent vacancies and vacancy defects.

| | Inherent vacancy | Vacancy defect |
|---|---|---|
| Free energy | enthalpy-driven | entropy-driven |
| Total number | nearly fixed | adjustable |
| Density (atoms cm$^{-3}$) | ~ $10^{22}$ | << $10^{20}$ |
| Site | atomic site at the lattice | |
| Local density of electron | electron-poor | |
| Chemical states | adjustable | |

In addition to the hexagonal phase, there are reports for additional phase of Sb$_2$Te$_3$, known as cubic phase. Here, the cation sites are randomly occupied by Sb and inherent vacancies at a ratio of 2:1, whereas the anion site is occupied by Te, as shown in structures reproduced by DFT calculations with a localized CD in the right panel in Figure 1a and transmission electron microscopy (TEM) images in Figure S4 in the Supporting Information. In this case, the inherent vacancies are on the alignments, identified by the results of the CDD (orange) and CD (grey). Notably, the cubic structure relatively relieves the electrostatic repulsion by increasing the volume, as indicated by the bond length between two Te (Te-Te) in the cubic phase (4.2 Å) compared with that in the hexagonal phase (3.7 Å). Thermodynamic studies such as differential scanning calorimetry prove that hexagonal and cubic phase is stable and metastable state, respectively. Specifically, cubic phase becomes metastable phase under the cost of the enthalpic loss due to electrostatic repulsion with inherent vacancies but profit of the entropic gain from the randomly distributed inherent vacancies. [23,24]

The site-switching of the inherent vacancies in the phases redistribute the excess electrons of Te as well as the chemical states of the elements in Sb$_2$Te$_3$. To clarify them, the chemical states were compared using the binding energies of core-level electrons obtained by X-ray photoelectron spectroscopy (XPS), as shown in Figure 1b. Te exhibits an identical symmetric single peak with the 2 δ − chemical state in the hexagonal phase. Further, an identical single peak is observed in the cubic phases, within the SET and RESET states, which represent low- and high-resistance states, respectively. On the other hand, the evolution of the chemical states of Sb is clearly observed depending on the phases. Specifically, the chemical state of Sb in the hexagonal phase is shown by a single spectrum of 3 δ +, while those in cubic phases can be deconvoluted to multiple spectra from 3 δ + to 1 δ + states. The fact that only one of the two elements in the compounds changes its chemical state without any stoichiometric modulation is atypical because the summation of the total evolution of the chemical states should be zero in a single compound. The uneven evolution can be explained by considering the chemical states of the numerous inherent vacancies occupying the atomic sites (1/6 of the total atomic sites), *i.e.*, the inherent vacancies have chemical states of 0 δ + in the hexagonal phase; while they have states of 0 δ +, 1 δ +, and 2 δ + in the cubic phases. As shown in Figure 1c, the chemical states of the inherent vacancies are consistent with the results of the DFT calculations, *i.e.*, the charge at the inherent vacancy sites of the hexagonal phase exceeds that of the cubic phase obtained by the spatial integration of the CD ($\int^r charge\ density,\ dV$). The direct observation of the evolution of the chemical states of the inherent vacancies makes experimental studies feasible to explore the kinetic behaviors of the inherent vacancies, as discussed below.

The evolution of the chemical states of inherent vacancies was verified to reduce the oxidation number during annealing, *i.e.*, from a few δ + to 0 δ +, corresponding to the phase change from the RESET to SET states, as shown in Figure 1d. The chemical states in $Sb_2Te_3$ gradually change during annealing up to 330 °C. In the case of the crystallization of GST (AIST) as a representative of the chalcogenide (pnictogenide)-dominant PCMs, Ge (In) drastically changes its chemical states near 180 °C, corresponding to the transition of the local environments from tetrahedral to octahedral. The evolution of the chemical states of the inherent vacancies in $Sb_2Te_3$ is different from those in GST and AIST. This can be explained by the unusual kinetic behaviors of the inherent vacancies, as discussed below, in comparison with the transition of the local environments in GST and AIST. When a certain number of inherent vacancies accumulate through site-switching with Sb to form a QL, atomic planes spontaneously glide to relieve the electrostatic repulsion, as shown by the red asterisk in Figure 1a (specific schematics are shown in Figure S6 in the Supporting Information). Because gliding is a collective motion of atoms, there are many possible cases, resulting in a divergence of the magnitude of the energy barrier for additional site-switching. Therefore, the evolution is continual up to a relatively high temperature (330 °C).

The changes in electrical resistances and structures that occur owing to the site-switching of inherent vacancies were further investigated, and the results are shown in Figure 1e,f. While drastic decreases in the electrical resistance during annealing are typically observed in GST and AIST, those are not found in $Sb_2Te_3$. The drastic changes of GST and AIST have been successfully explained by the transition of the local environment of certain elements at early reports such as the umbrella flip, bond-interchange, and dam-gate models. [13,14,25] On the other hand, the gradual change of $Sb_2Te_3$ without drastic decrease can be explained with the characteristic behavior of site-switching of inherent vacancies with diverged magnitude of the energy barrier. The lattice parameters along the c-axis direction also smoothly contract, as shown in Figure 1f. From the figure, the X-ray diffraction (XRD) peak position in the direction of $Sb_2Te_3$ (0018) smoothly increases from 51.8° to 54.3°, indicating a decrease of 3.3% in the length scale, where each peak position agrees with the DFT calculation. The electrical and structural results agree with the gradual evolution, *i.e.*, the specific kinetic behavior of the inherent vacancies, corresponding to a second-order transition, is different from that of the elements in GST and AIST, corresponding to a first-order transition.

Phase transition from the stable hexagonal to metastable cubic phases increases the oxidation number of the inherent vacancies from 0 δ + to a few δ +. Some experimental studies have shown that it requires significantly more power and time, compared with the phase change at the metastable states between the SET and RESET states, because the stable hexagonal phase is energetically too stable to change phases. The results could be explained using the chemical states of the inherent vacancies. As sites of inherent vacancies in 0 δ + chemical states in the hexagonal phase are out of the alignments and occupy relatively squeezed space than those in the cubic phase, the site-switching of the inherent vacancies intensifies the electrostatic repulsion, *i.e.*, higher energy barriers are formed against increasing oxidation number of the inherent vacancies from zero (phase-transition from stable to metastable states) in comparison with that against changing the number in the positively charged inherent vacancies (phase change at the metastable states). Therefore, much more energy is needed to drive an increase from the 0 δ + to the positively charged states of the inherent vacancies, *i.e.*, complete melt quenching.

The study on the kinetic behavior of inherent vacancies could be extended from $Sb_2Te_3$ to the superlattices composed of 2D and 3D sublayers, $Sb_2Te_3$ and GeTe. Herein, in order to resolve the long-standing controversies on the mechanism, phases in the metastable and stable states of the superlattices were investigated with characterizing bond types based on the kinetic behavior of the inherent vacancies at the interfaces. [26–28]

Two different types of interfaces are expected between the 3D (GeTe) and 2D ($Sb_2Te_3$) sublayers *i.e.*, GeTe sublayers prefer direct chemical bonds at the interfaces; while $Sb_2Te_3$ sublayers prefer indirect vdW bonds. As shown in Figure 2, two different types of the interfaces between sublayers are directly observed with intensity profiling STEM methods. The STEM image shows that Ge exhibits a weaker signal compared with Sb and Te owing to its low effective charge (Z*). The structure at the interface also is corroborated using DFT calculation as shown in the lower panels. CDD results differentiate the type of bonding, *i.e.*, whether direct or indirect chemical bonds are formed over the interfaces between adjacent Te layers, as displayed in the dashed orange box. Two types of interfaces are observed with Te-Te chemisorption and Te-Te physisorption, shown by the red and blue tilde symbols, respectively. Comparing the structures of the two different types of the interfaces, the two interfaces have distinct structure, such as the Te–Te distances at each interface and atomic stacking sequence between the sublayers. The Te–Te distances over the interfaces are as short as 10% in the Te-Te chemisorption interface (3.28 Å) compared with that in the Te-Te physisorption interface (3.61 Å). Moreover, at the Te-Te chemisorption interface, cations occupy the same atomic (*a-* and *a-*) sites over the interfaces, whereas, at the Te-Te physisorption interface, they occupy different atomic (*a- and b-*) sites.

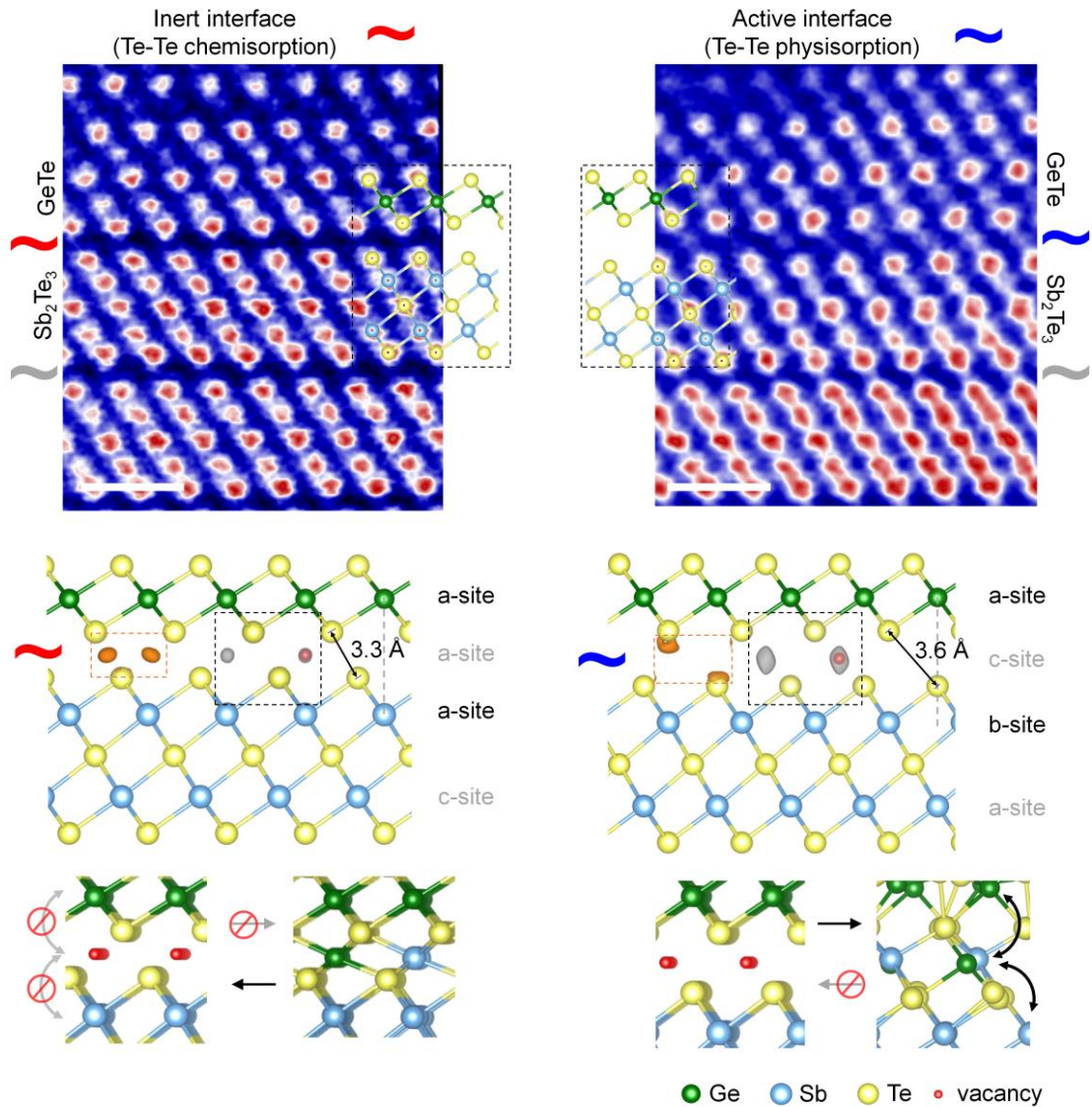

**Figure 2. Two different types of interfaces between the GeTe and Sb$_2$Te$_3$ sublayers**. Te-Te chemisorption and Te-Te physisorption interfaces between the GeTe and Sb$_2$Te$_3$ sublayers. Scale bar is 1-nm long.

To study the kinetics of the inherent vacancies depending on the interfaces, the sites of inherent vacancies (red balls) are defined with CD, which is the same method used above with Sb$_2$Te$_3$. Upon site-switching between inherent vacancies and cations (Ge and Sb), inherent vacancies at the Te-Te chemisorption interfaces failed to relax and go back to their original sites while those at the Te-Te physisorption interfaces are successfully stabilized, irrespective of the type of site-switched cations (Ge and Sb), relaxing procedures, and size of the cells, as displayed in Figure S7 in the Supporting Information. The intensified repulsion upon site-switching of the inherent vacancies at the Te-Te chemisorption interface is the origin of the failure of the relaxation because the sites of inherent vacancies are out of alignments in $p$-orbital electrons and the space is rather squeezed, which is similar to the difficulty for the phase transition of $Sb_2Te_3$ from stable hexagonal to metastable cubic phase. However, in the case of the Te-Te physisorption interface, the cations could occupy their sites with sufficient space and form bonds with Te under the alignment even after site-switching with the inherent vacancies. Therefore, the kinetic behaviors of the inherent vacancies characterize the Te-Te chemisorption and Te-Te physisorption interfaces as an inert and active interface, respectively. Since the site switching at the active (Te-Te physisorption) interface is favorable owing to the configurational entropy, cation diffusion at the interface is inevitable during the fabrication of the superlattices. The diffusion agrees with the intensity profiling in the STEM results: the signal in intensity profiling caused by cation diffusion is suppressed at the inert (Te-Te chemisorption) interfaces; while that of non-negligible atomic periodicity owing to the diffusion is probed at the active interfaces. [29]

Based on the obtained kinetic behavior at the two different interfaces, we've further studied phases at the superlattices to clarify the metastable and stable states. As the memory performance is maximized when additional Te is introduced in the GeTe sublayers, specifically in the composition of $[Ge_1Te_2/(Sb_2Te_3)_4]_n$, the superlattice was fabricated in the composition with molecular beam epitaxy (MBE) and reproduced with DFT calculation as shown in Figure 3. The superlattice grown at 160 °C consists of a block with three atomic sublayers sandwiched between two blocks with five atomic sublayers (5/3/5 layers), as shown in the left panel. Because Ge exhibits a far weaker signal than Sb and Te, three atomic sublayers are denoted as the Te-Ge-Te (TGT) stack, whereas the other five atomic sublayers are denoted as $Sb_2Te_3$. The classification denoting three atomic sublayers as TGT rather than Te–Sb–Te is consistent with the previous TEM results that 5/3/5 layers are not observed in Sb–Te compounds but 5/5/5, 5/7/3, and 5/3/7 atomic planes, which agrees well with the electron counting model, as displayed in Figure **S3** in the Supporting Information.[17] Although the stacking sequence is identical to that of the ferro phase, a considerable number of inherent vacancies are implemented at the cationic sites with moderate intermixing between Ge and Sb. Based on these structural characteristics, we represent the phase as an extended ferro (Xferro) phase containing the inert interface between the TGT and upper $Sb_2Te_3$ sublayers (red tilde). As discussed above, site switching of inherent vacancies between Ge and Sb could occur inside the Xferro phase (blue tilde) rather than at the inert interface (red tilde) between the Xferro phase and upper $Sb_2Te_3$.

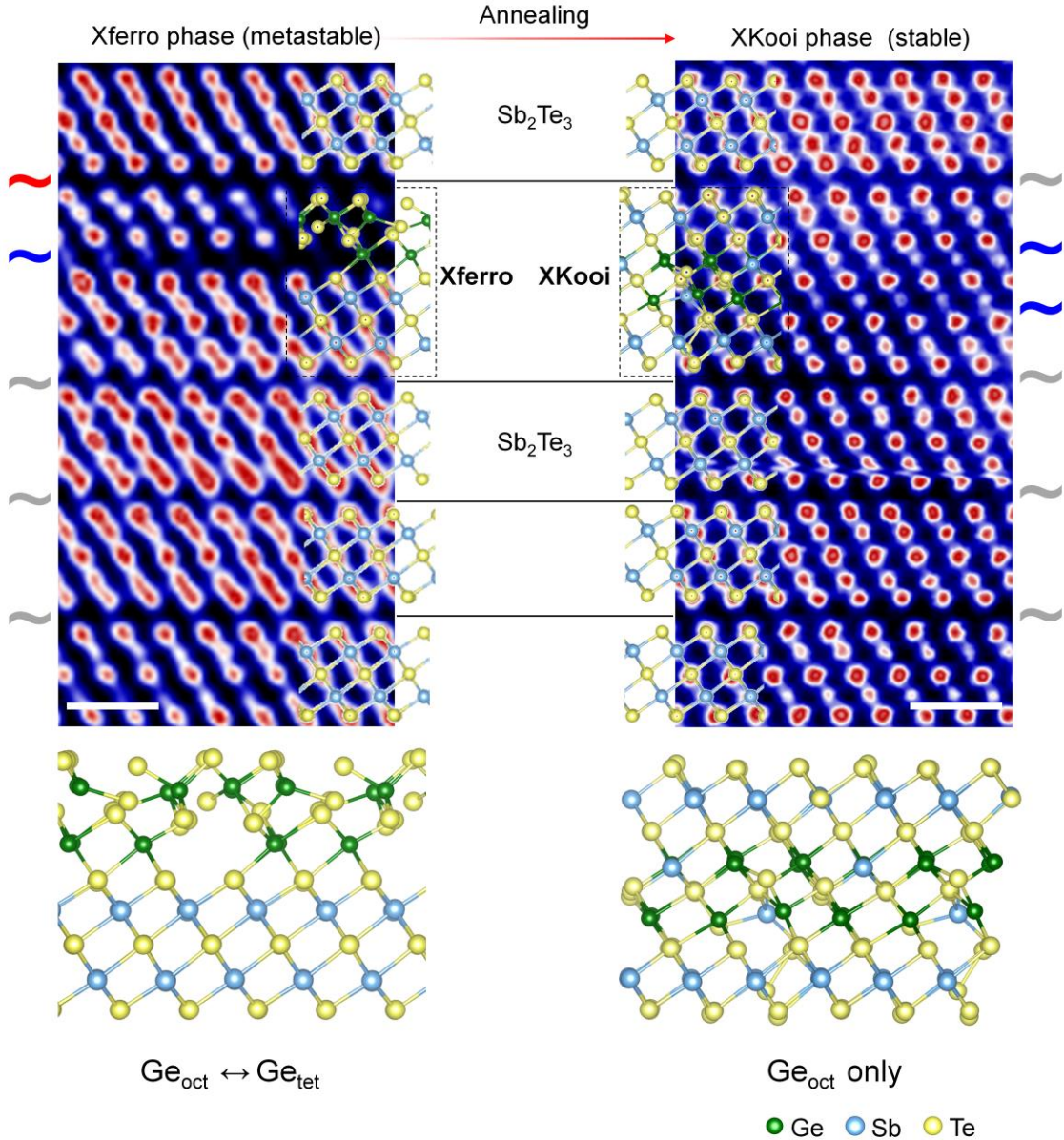

**Figure 3. [(Ge₁Te₂) / (Sb₂Te₃)₄]ₙ superlattice**. Xferro and XKooi phase. Scale bar is 1-nm long.

Additionally, annealing the superlattice with the Xferro phase at 300 °C for 30 min, site-switching between Ge (Sb) and inherent vacancies follows, as shown on the right panel in Figure 3. According to the intensity profiling results, cation sites far from the vdW layer are occupied by Ge, whereas those near the vdW layer are occupied by Sb, which is also consistent with the experimental and calculation reports that the vdW layer by the –Sb–Te termination is considerably more stable than that by the –Ge–Te termination.[30,31] Although the stacking sequence is identical to that of the Kooi phase, a considerable number of inherent vacancies are implemented at cationic sites with moderate intermixing between Ge and Sb. Similar to the Xferro phase, the phase is represented as an extended Kooi (XKooi) phase.[15,32] Upon site switching between the inherent vacancies and cations (Ge and Sb), the transition in the local environment represents apparent changes depending on the phases. Local environments in the XKooi phase remain octahedron irrespective of the site switching. On the other hand, the local environments of Ge considerably change in the Xferro phase, while the local environments of Sb also rarely change, as displayed in Figure 4a. When nearest cation sites from Ge are occupied by inherent vacancies rather than by Sb, the local environments of Ge spontaneously change from octahedron to tetrahedron, as indicated by the pink arrows. It is also consistent with the results of *in situ* XPS, where the evolution in the chemical states of Ge are observed between SET and RESET phases of the superlattices.[33] It is a notable result that spontaneous transition from octahedron to tetrahedron has been achieved during the structural relaxation of the superlattices.

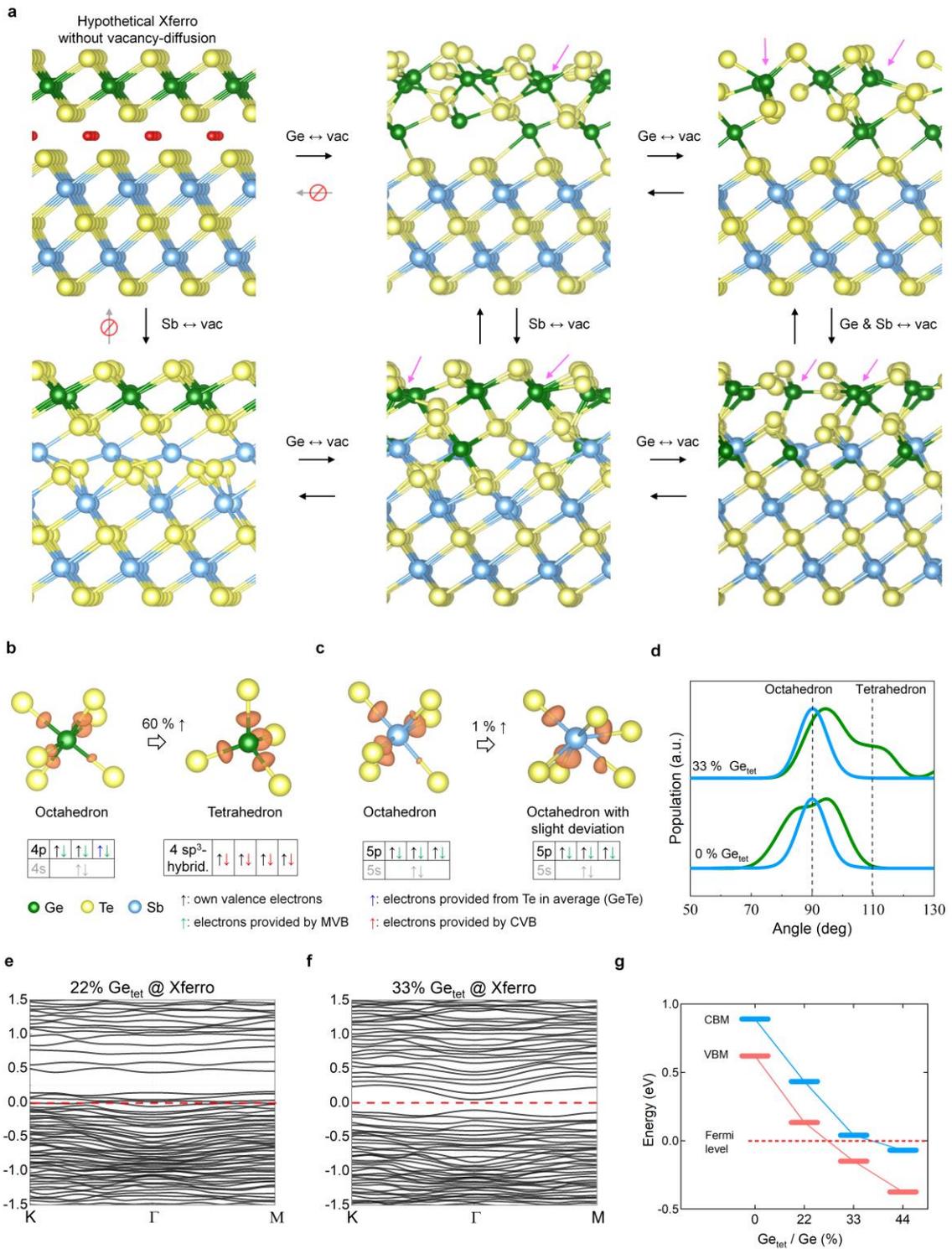

**Figure 4. Transition in local environment at Xferro phase**. (a) Site switching at the Xferro phase (b and c) Characterization of chemical bond depending on the local structures of Ge and Sb. The electron density of Ge and Sb is 0.006 and 0.004 e bohr$^{-3}$, respectively. (d) Angle distribution function of atoms (Ge and Sb) of various cells. (e and f) Band structures of various Xferro cells depending on the local environments of Ge. (g) Band diagrams depending on the local environment of Ge.

Depending on the local environments of Ge, the number of electrons per bond between the nearest atoms represent far different features, as identified by the calculation on the electron distribution at the chemical bonds via the DDEC6 methods, specialized in calculating bond order. The average number of electrons per bond of Ge in the Xferro phases is calculated as 1.016 (1.624) at the octahedron (tetrahedron), corresponded to the value of MVB (CVB). [30,34,35] Those results agree well with the heuristic electron counting model as shown in Figure 4b and Figure S8 in the Supporting Information. By completing the valence shells with different types of bond, there are two independent local environments in Ge, corresponding to the $p^3$ (octahedron) and $sp^3$ (tetrahedron) orbital configurations. Meanwhile, the local environment of Sb only maintains octahedron, where the calculated number ranges from 1.036 to 1.048, also corresponding to the value of MVB. Those results also agree with the heuristic model, where the valence shell of Sb is completed via six MVBs, providing three electrons ($p^{3+3=6}$). As Sb already has three $p$ electrons in the valence shell, it cannot change its local environments into tetrahedron.

This study reproduced cells with site-switching at the Xferro phases to verify the band structures depending on the local environments of Ge. During the site-switching of the inherent vacancies at Xferro phases, the transition of the local environments of Ge from octahedron to tetrahedron is found in angle distributions, as shown in Figure 4d. Moreover, changes in the band structures between metallic and semiconducting behaviors are also confirmed, as shown in Figure 4e, f. The Fermi level is shifted downward from the conduction band to the valence band, while the energy gap is maintained as the content increases, as displayed in Figure 4g. These results directly suggest that about 30% of the content of Ge in tetrahedron drastically changes the type of the bond which is consistent with the experimentally estimated value of the contents (30%) in the alloys via structural analysis.[36] As the conversion in bond type between MVB and CVB could be triggered by the collapse in middle range order, it is proposed that the transition in the local environments of Ge in part from octahedron to tetrahedron is responsible for the conversion from MVB to CVB at the Xferro superlattices. Further evidences for criteria comparing MVB and CVB materials such as optical spectra in precedent report which first discussed the bonding mechanism are provided Supplementary Figure S9. Referring to the recent studies which successfully account for the clear contrast in the electrical resistance by more than two orders of magnitude upon phase-change with the conversion in bond types, the transition of Ge is responsible for the contrast in the electrical resistances of the superlattices.

The phase-change mechanism based on the transition of Ge suggested at the Xferro superlattices is clearly differentiated from the previous reports, which are based on four superlattice models, namely the ferro, Kooi, inverted Petrov, and Petrov phases. Because these four phases exhibit distinct *abc*-atomic site occupancies and exclude tetrahedral local environments, the reported phase-change mechanisms in their structure models inevitably include the gliding of the atomic plane rather than a transition in the local environment of elements.[31,37,38] However, because gliding is a collective motion of atoms considered as the origin of the second-order transition, it contradicts the superior cyclability ($\sim 10^9$) and nanosecond-scale ultrafast phase-change.[39,40] By contrast, the transition of Ge at the Xferro superlattices successfully accounts for the phase change without the gliding process but transition in the local environment. Furthermore, in previous reports of the Ge-switch models, the inverted Petrov phase composed of homopolar bonds of Ge (Ge-Ge) was considered as the HRS superlattices, which conflicts with the earlier bond topology models (or ring statistics model), which accounts for the inferior memory performances of GeTe alloys compared to those of GST alloys with the presence of Ge-Ge bonding.[41] According to the model, the unintended Ge–Ge bonding in amorphous GeTe fosters phase-separation which is suppressed in the case of GST, which degrades performances such as cyclabilities. Therefore, Ge–Ge bonding in the mechanism of superlattices is hard to be accepted because the cyclabilities of superlattices (~ $10^9$) surpass those of GST (~ $10^6$) and GeTe (~ $10^4$) alloys. Finally, the controversy is resolved as the Ge homopolar bond (Ge–Ge) is excluded even in Ge with the tetrahedral local environment at the Xferro phase.

The transition of Ge at the Xferro phases extends the comprehensive understanding of the phase-change mechanism ranging from the Ge-Sb-Te alloys to superlattices. Previous reports of GST alloys successfully verify that the transition of Ge only occurs in the metastable cubic phases, not in the stable hexagonal phases. Since the transition of the superlattices only occurs in the Xferro, not in the XKooi phases, the Xferro and XKooi phases can be corresponded to the metastable and stable states, respectively, which is also consistent with the results that XKooi phases are also commonly found in GST alloys of the stable hexagonal phase rather than the metastable cubic phase. Further, the phase-transition from XFerro to XKooi phases through the additional annealing process agrees with the phase-transition from the metastable cubic to stable hexagonal phases in the case of GST alloys.[42] We have to refer two different precedent studies of phase-change mechanism of GST alloys whether transition of local environments of Ge upon phase-change is between 'octahedron and tetrahedron' or 'octahedron and defective octahedron', which are specifically discussed in Supplementary Figure S10. [13,43]

In the case of other chalcogenide PCMs, TM-doped $Sb_2Te_3$, there are also prominent evolutions in the chemical states of inherent vacancies, such as $Ti_{0.32}Sb_2Te_3$ and $Ag_{0.17}Sb_{1.83}Te_3$. Similar to $Sb_2Te_3$, the XPS peak of Sb $3d_{5/2}$ shows asymmetric spectra of Sb $3d_{5/2}$, deconvoluted with several chemical states, e.g. 3 δ +, 2 δ +, or 1 δ+, and symmetric spectra of Te $3d_{5/2}$, identified with a chemical state of a 2 δ -, as presented in Figure S11 in the Supporting Information. When comparing the phase change of the TM-doped $Sb_2Te_3$ with that of the undoped $Sb_2Te_3$, the evolution of the chemical states of Sb in the TM-doped $Sb_2Te_3$ is considerably suppressed, which indicates that doping TM confines the evolution of chemical states of inherent vacancies. The suppressed diversity in chemical states upon site switching between Sb and inherent vacancies implies that the distribution of the formation energy and activation energy for site switching is restrained, which agrees with the octahedral motif that TM doping reduces stochasticity in octahedral local structures even in the amorphous phase to enhance memory performance.[44]

**Conclusions**

Direct observation on the evolution of the chemical states of the inherent vacancies during the phase-transition reveals their kinetic behavior as a second order transition owing to the spontaneous gliding of atomic plane when inherent vacancies are accumulated. The study on the kinetics is extended to superlattices that the site-switching of the inherent vacancies is prevented (allowed) at the inert (active) interfaces between 2D and 3D sublayers. The distinct behavior of the inherent vacancies depending on the interfaces classifies the phases of the superlattices into Xferro (metastable) and XKooi (stable) phases, where the conversion in bond types between the MVB and CVB is reproduced only at the Xferro phases, not in XKooi states. The conversion between MVB and CVB in metastable superlattices composed of 2D and 3D sublayers suggest the phase-change mechanism, which resolves long-standing controversies over a decade.

Dynamical behaviors of the compounds are reformed for use as a memory via extending category of crystal structures from stable to metastable states driven by redistribution of the inherent vacancies. Inherent vacancies change their atomic sites based on the competition between enthalpic and entropic gain that minimizes the electrostatic repulsion at their sites and maximizes randomness in atomic configurations, respectively. The direct observation on the evolution in chemical states of inherent vacancies would extend the experimental limits by combining with the state-of-art techniques on fragility, time-resolved ultrafast transition, Peierls distortion, and coherent atomic displacements.[15–18,45–50] Those interdisciplinary convergence would extend the understanding on the phase-change operations such as kinetic physical factors of inherent vacancies for phase-transition speed and activation energy. As the materials discussed also have considerable interest as topological insulators and Weyl semimetals, understanding the kinetics of their inherent vacancies would inspire studies on the change in spin textures.[51] This study, which is distinct from the other conventional approaches that directly explore the kinetics and chemical states of vacancies, can contribute to realizing vacancy engineering in various functional materials.[52–56]

**Methods**

**Sample Preparation** The films were fabricated using the MBE and sputtering methods. In the MBE methods, the sample was grown by a two-step process on a Si (111) substrate at a base pressure of $9 \times 10^{-9}$ Torr after vacuum annealing at 600 °C for surface cleaning. First, $Sb_2Te_3$ was deposited at room temperature as a seed layer through a thermal effusion cell (1.6 Å min$^{-1}$ for Sb and 3 Å min$^{-1}$ for Te) and post-annealed at 180 °C for 30 min. In the second step, $Sb_2Te_3$ and GeTe were deposited at 160 °C. At this stage, Ge was also provided via effusion at a rate of 1.0 Å min$^{-1}$. For the sputtering methods, $Ar^+$ ion beam sputtering was used targeting the desired composition at a base pressure of $3 \times 10^{-8}$ Torr and a working pressure of $7.8 \times 10^{-5}$ Torr. The compositions were confirmed through *in situ* XPS.

**Transmission Electron Microscopy (TEM)** Cross-sectional specimens for TEM characterization were prepared using a focused ion beam (FIB, NX2000; Hitachi Inc., Japan) instrument with low-accelerated $Ga^+$ ions. High resolution (HR)-TEM and high angle annular dark field scanning electron microscopy (HAADF-STEM) were performed using JEM3010, JEM2100F, and ARM200F.

**DFT** The electronic structures of the $Sb_2Te_3$ and $[GeTe_2/(Sb_2Te_3)_4]_n$ superlattices were calculated using DFT with the Vienna Ab initio Simulation Package (VASP)[57]. All calculations were implemented at the generalized gradient approximation level, using the Perdew–Burke–Ernzerhof (PBE) functional[58]. The Grimme's D3 dispersion correction term[59] was added to capture long-range interactions during structure optimization. The Brillouin zone was sampled with 11 × 11 × 11 (21 × 21 × 5) k-points, the kinetic energy cutoff for the plane wave was set to 500 eV (400 eV), and the convergence criterion for the self-consistent loop was 1E-6 eV (1E-4 eV) for optimization (single-point calculation) until the difference in force between steps was less than 0.01 eV Å$^{-1}$. The unit cell of $Sb_2Te_3$ contained three quintuple layers, with the Sb and Te layers stacked alternately with one atom per layer. The positions of the atoms, cell shapes, and cell volumes were determined without limitations.

The $Sb_2Te_3$ supercell and $[GeTe_2/(Sb_2Te_3)_4]_n$ superlattices were relaxed with 2 × 2 × 2 Brillouin zone sampling and gamma point sampling, respectively. By fabricating several 3 × 3 × 1 supercells of $Sb_2Te_3$, representative structures were created in which Sb and vacancies were mixed in a ratio of 2:1 in all layers except for the Te layers. The exact location of the vacancy was determined by the CD and monitored during the process of site switching and optimization. Subsequently, the CD near the vacancy site was numerically integrated. The $[GeTe_2/(Sb_2Te_3)_4]_n$ superlattices were constructed in the same manner. The vacancies, Sb, and Ge were mixed at the interface between the $GeTe_2$ and $Sb_2Te_3$. The bond order was calculated using the density-derived electrostatic and chemical (DDEC6) approach.[60] The number of electrons is twice as much as the bond order. The band structure calculations were performed along the symmetry points (K, Γ, M) in the reciprocal space.

**X-ray Photoelectron Spectroscopy (XPS)** *In situ* XPS profiles of the core levels were measured using a PHI 5000 VersaProbe made by ULVAC-PHI, where the base pressure was $5.0^{-10}$ Torr. The XPS spectra were recorded using monochromatic Kα radiation with an analyzer pass energy of 23.5 and 11.75 eV, providing an overall experimental resolution of 200 meV. The background signal of each spectrum was subtracted using the Shirley method. To confirm whether the obtained data represent bulk or surface properties, the samples were measured by varying the sampling depth by tilting the samples from 45° to 80° with respect to the detector.

The results revealed that the obtained data represent the bulk properties without the oxidized state, where the crystallinity of the crystalline samples is sufficient.

**Resistance Dependence on Temperature Under Annealing and Cooling (R-T)** The measurements were performed after several cycles of alternating vacuum pumping and $N_2$ purging. A constant current mode of 5 mA was used for the two-point probe. The sample width, length, and thickness were 1, 2, and 300 nm, respectively.

**Supporting Information**

Supporting Information is available at htttps://pubs.acs.org/doi/XX.

Figure S1-S13. TEM images, DFT results, electron counting model, Schematics, and XPS results.

**Acknowledgements**

$^{\triangledown}$Dasol Kim, Youngsam Kim, Jin-Su Oh, Changwoo Lee, and Hyeonwook Lim contributed equally to this work. This work was supported by the National Research Foundation of Korea (NRF) grants funded by the Government of Korea (MSIP) (No. 2021M3H4A1A03052566 and No. 2020M3F3A2A0108232413), the Ministry of Trade, Industry & Energy (MOTIE) of Korea (Project No. 10080625), and the Korea Semiconductor Research Consortium (KSRC) through a project for developing source technologies for future semiconductor devices. Y.K. and E.S. acknowledge the support from NRF-2020R1A2C2007468.

**Conflict of Interest**

The authors declare that they have no competing financial interests.